\documentclass[prd,superscriptaddress,showpacs,amssymb,amsmath,amsfonts,aps,altaffilletter,nofootinbib,letterpaper,twocolumn]{revtex4-1}

\usepackage{color}
\usepackage{amsfonts}
\usepackage{graphicx}
\usepackage{fancyhdr}
\usepackage{float}
\usepackage{ulem}
\usepackage[raggedright]{subfigure}
\usepackage{acronym}
\usepackage{multirow}
\usepackage{array}
\usepackage{hyperref}
\usepackage{appendix}

\makeatletter
\makeatletter
\def\@fnsymbol#1{\ifcase#1\or * \or  $+$ \or  \$ \or \#  \or \dag \or \ddag \or
$\mathsection$ \or $ \mathparagraph$ \or $\|$  \or \textordfeminine \or \textbul
let   
\or ** \or $++$ \or  \$\$ \or \#\#  \or \dag\dag \or \ddag\ddag \or
$\mathsection\mathsection$ \or $ \mathparagraph\mathparagraph$ \or $\|\|$  \or 
\textordfeminine\textordfeminine \or \textbullet \textbullet \or *** \or $+++$ 
\or  \$\$\$ \or \#\#  \or \dag\dag \or \ddag\ddag \or
$\mathsection \mathsection\mathsection$ \or $ \mathparagraph 
\mathparagraph\mathparagraph$ \or $\|\|\|$  \or 
\textordfeminine\textordfeminine\textordfeminine \or 
\textbullet\textbullet\textbullet \or \else \@ctrerr\fi}
\makeatother

\newcommand\fake[1]{\textcolor{red}{#1}}

\renewcommand{\today}{\number\day\space\ifcase\month\or
  January\or February\or March\or April\or May\or June\or
  July\or August\or September\or October\or November\or December\fi
  \space\number\year}

\begin{document}

\title{Search for gravitational wave ringdowns from perturbed intermediate mass black holes
in LIGO-Virgo data from 2005-2010} 

\author{%
J.~Aasi$^{1}$,
B.~P.~Abbott$^{1}$,
R.~Abbott$^{1}$,
T.~Abbott$^{2}$,
M.~R.~Abernathy$^{1}$,
F.~Acernese$^{3,4}$,
K.~Ackley$^{5}$,
C.~Adams$^{6}$,
T.~Adams$^{7}$,
P.~Addesso$^{4}$
% ...
}\noaffiliation

\affiliation {LIGO, California Institute of Technology, Pasadena, CA 91125, USA }
\affiliation {Louisiana State University, Baton Rouge, LA 70803, USA }
\affiliation {INFN, Sezione di Napoli, Complesso Universitario di Monte S.Angelo, I-80126 Napoli, Italy }
\affiliation {Universit\`a di Salerno, Fisciano, I-84084 Salerno, Italy }
\affiliation {University of Florida, Gainesville, FL 32611, USA }
\affiliation {LIGO Livingston Observatory, Livingston, LA 70754, USA }
\affiliation {Cardiff University, Cardiff, CF24 3AA, United Kingdom }
\affiliation {Albert-Einstein-Institut, Max-Planck-Institut f\"ur Gravitationsphysik, D-30167 Hannover, Germany }
\affiliation {Nikhef, Science Park, 1098 XG Amsterdam, The Netherlands }
\affiliation {LIGO, Massachusetts Institute of Technology, Cambridge, MA 02139, USA }
\affiliation {Instituto Nacional de Pesquisas Espaciais, 12227-010 - S\~{a}o Jos\'{e} dos Campos, SP, Brazil }
\affiliation {Inter-University Centre for Astronomy and Astrophysics, Pune - 411007, India }
\affiliation {International Centre for Theoretical Sciences, Tata Institute of Fundamental Research, Bangalore 560012, India. }
\affiliation {Syracuse University, Syracuse, NY 13244, USA }
\affiliation {University of Wisconsin--Milwaukee, Milwaukee, WI 53201, USA }
\affiliation {Leibniz Universit\"at Hannover, D-30167 Hannover, Germany }
\affiliation {INFN, Sezione di Pisa, I-56127 Pisa, Italy }
\affiliation {Universit\`a di Siena, I-53100 Siena, Italy }
\affiliation {Stanford University, Stanford, CA 94305, USA }
\affiliation {The University of Mississippi, University, MS 38677, USA }
\affiliation {California State University Fullerton, Fullerton, CA 92831, USA }
\affiliation {INFN, Sezione di Roma, I-00185 Roma, Italy }
\affiliation {University of Birmingham, Birmingham, B15 2TT, United Kingdom }
\affiliation {Albert-Einstein-Institut, Max-Planck-Institut f\"ur Gravitationsphysik, D-14476 Golm, Germany }
\affiliation {Montana State University, Bozeman, MT 59717, USA }
\affiliation {European Gravitational Observatory (EGO), I-56021 Cascina, Pisa, Italy }
\affiliation {LIGO Hanford Observatory, Richland, WA 99352, USA }
\affiliation {SUPA, University of Glasgow, Glasgow, G12 8QQ, United Kingdom }
\affiliation {APC, AstroParticule et Cosmologie, Universit\'e Paris Diderot, CNRS/IN2P3, CEA/Irfu, Observatoire de Paris, Sorbonne Paris Cit\'e, 10, rue Alice Domon et L\'eonie Duquet, F-75205 Paris Cedex 13, France }
\affiliation {Columbia University, New York, NY 10027, USA }
\affiliation {Universit\`a di Pisa, I-56127 Pisa, Italy }
\affiliation {CAMK-PAN, 00-716 Warsaw, Poland }
\affiliation {Astronomical Observatory Warsaw University, 00-478 Warsaw, Poland }
\affiliation {The University of Texas at Brownsville, Brownsville, TX 78520, USA }
\affiliation {INFN, Sezione di Genova, I-16146 Genova, Italy }
\affiliation {Universit\`a degli Studi di Genova, I-16146 Genova, Italy }
\affiliation {San Jose State University, San Jose, CA 95192, USA }
\affiliation {Faculty of Physics, Lomonosov Moscow State University, Moscow 119991, Russia }
\affiliation {LAL, Universit\'e Paris-Sud, IN2P3/CNRS, F-91898 Orsay, France }
\affiliation {NASA/Goddard Space Flight Center, Greenbelt, MD 20771, USA }
\affiliation {University of Western Australia, Crawley, WA 6009, Australia }
\affiliation {Department of Astrophysics/IMAPP, Radboud University Nijmegen, P.O. Box 9010, 6500 GL Nijmegen, The Netherlands }
\affiliation {Universit\'e Nice-Sophia-Antipolis, CNRS, Observatoire de la C\^ote d'Azur, F-06304 Nice, France }
\affiliation {Institut de Physique de Rennes, CNRS, Universit\'e de Rennes 1, F-35042 Rennes, France }
\affiliation {Laboratoire d'Annecy-le-Vieux de Physique des Particules (LAPP), Universit\'e de Savoie, CNRS/IN2P3, F-74941 Annecy-le-Vieux, France }
\affiliation {Washington State University, Pullman, WA 99164, USA }
\affiliation {INFN, Sezione di Perugia, I-06123 Perugia, Italy }
\affiliation {INFN, Sezione di Firenze, I-50019 Sesto Fiorentino, Firenze, Italy }
\affiliation {Universit\`a degli Studi di Urbino 'Carlo Bo', I-61029 Urbino, Italy }
\affiliation {University of Oregon, Eugene, OR 97403, USA }
\affiliation {Laboratoire Kastler Brossel, ENS, CNRS, UPMC, Universit\'e Pierre et Marie Curie, F-75005 Paris, France }
\affiliation {VU University Amsterdam, 1081 HV Amsterdam, The Netherlands }
\affiliation {University of Maryland, College Park, MD 20742, USA }
\affiliation {University of Massachusetts Amherst, Amherst, MA 01003, USA }
\affiliation {Laboratoire des Mat\'eriaux Avanc\'es (LMA), IN2P3/CNRS, Universit\'e de Lyon, F-69622 Villeurbanne, Lyon, France }
\affiliation {Universitat de les Illes Balears, E-07122 Palma de Mallorca, Spain }
\affiliation {Universit\`a di Napoli 'Federico II', Complesso Universitario di Monte S.Angelo, I-80126 Napoli, Italy }
\affiliation {Canadian Institute for Theoretical Astrophysics, University of Toronto, Toronto, Ontario, M5S 3H8, Canada }
\affiliation {Tsinghua University, Beijing 100084, China }
\affiliation {University of Michigan, Ann Arbor, MI 48109, USA }
\affiliation {Rochester Institute of Technology, Rochester, NY 14623, USA }
\affiliation {INFN, Sezione di Roma Tor Vergata, I-00133 Roma, Italy }
\affiliation {National Tsing Hua University, Hsinchu Taiwan 300 }
\affiliation {Charles Sturt University, Wagga Wagga, NSW 2678, Australia }
\affiliation {Caltech-CaRT, Pasadena, CA 91125, USA }
\affiliation {Pusan National University, Busan 609-735, Korea }
\affiliation {Australian National University, Canberra, ACT 0200, Australia }
\affiliation {Carleton College, Northfield, MN 55057, USA }
\affiliation {INFN, Gran Sasso Science Institute, I-67100 L'Aquila, Italy }
\affiliation {Universit\`a di Roma Tor Vergata, I-00133 Roma, Italy }
\affiliation {Universit\`a di Roma 'La Sapienza', I-00185 Roma, Italy }
\affiliation {University of Brussels, Brussels 1050 Belgium }
\affiliation {Sonoma State University, Rohnert Park, CA 94928, USA }
\affiliation {Embry-Riddle Aeronautical University, Prescott, AZ 86301, USA }
\affiliation {The George Washington University, Washington, DC 20052, USA }
\affiliation {University of Cambridge, Cambridge, CB2 1TN, United Kingdom }
\affiliation {Northwestern University, Evanston, IL 60208, USA }
\affiliation {University of Minnesota, Minneapolis, MN 55455, USA }
\affiliation {The University of Sheffield, Sheffield S10 2TN, United Kingdom }
\affiliation {Wigner RCP, RMKI, H-1121 Budapest, Konkoly Thege Mikl\'os \'ut 29-33, Hungary }
\affiliation {University of Sannio at Benevento, I-82100 Benevento, Italy and INFN, Sezione di Genova, I-16146 Genova, Italy }
\affiliation {INFN, Gruppo Collegato di Trento, I-38050 Povo, Trento, Italy }
\affiliation {Universit\`a di Trento, I-38050 Povo, Trento, Italy }
\affiliation {Montclair State University, Montclair, NJ 07043, USA }
\affiliation {The Pennsylvania State University, University Park, PA 16802, USA }
\affiliation {MTA E\"otv\"os University, `Lendulet' A. R. G., Budapest 1117, Hungary }
\affiliation {Universit\`a di Perugia, I-06123 Perugia, Italy }
\affiliation {Accuray Inc., Sunnyvale, CA 94089, USA }
\affiliation {Rutherford Appleton Laboratory, HSIC, Chilton, Didcot, Oxon, OX11 0QX, United Kingdom }
\affiliation {Perimeter Institute for Theoretical Physics, Ontario, N2L 2Y5, Canada }
\affiliation {American University, Washington, DC 20016, USA }
\affiliation {University of Adelaide, Adelaide, SA 5005, Australia }
\affiliation {Raman Research Institute, Bangalore, Karnataka 560080, India }
\affiliation {Korea Institute of Science and Technology Information, Daejeon 305-806, Korea }
\affiliation {Bia{\l }ystok University, 15-424 Bia{\l }ystok, Poland }
\affiliation {University of Southampton, Southampton, SO17 1BJ, United Kingdom }
\affiliation {IISER-TVM, CET Campus, Trivandrum Kerala 695016, India }
\affiliation {Institute of Applied Physics, Nizhny Novgorod, 603950, Russia }
\affiliation {Seoul National University, Seoul 151-742, Korea }
\affiliation {Hanyang University, Seoul 133-791, Korea }
\affiliation {IM-PAN, 00-956 Warsaw, Poland }
\affiliation {NCBJ, 05-400 \'Swierk-Otwock, Poland }
\affiliation {Institute for Plasma Research, Bhat, Gandhinagar 382428, India }
\affiliation {The University of Melbourne, Parkville, VIC 3010, Australia }
\affiliation {INFN, Sezione di Padova, I-35131 Padova, Italy }
\affiliation {Monash University, Victoria 3800, Australia }
\affiliation {SUPA, University of Strathclyde, Glasgow, G1 1XQ, United Kingdom }
\affiliation {ESPCI, CNRS, F-75005 Paris, France }
\affiliation {Argentinian Gravitational Wave Group, Cordoba Cordoba 5000, Argentina }
\affiliation {Universit\`a di Camerino, Dipartimento di Fisica, I-62032 Camerino, Italy }
\affiliation {The University of Texas at Austin, Austin, TX 78712, USA }
\affiliation {Southern University and A\&M College, Baton Rouge, LA 70813, USA }
\affiliation {College of William and Mary, Williamsburg, VA 23187, USA }
\affiliation {IISER-Kolkata, Mohanpur, West Bengal 741252, India }
\affiliation {National Institute for Mathematical Sciences, Daejeon 305-390, Korea }
\affiliation {Hobart and William Smith Colleges, Geneva, NY 14456, USA }
\affiliation {RRCAT, Indore MP 452013, India }
\affiliation {Tata Institute for Fundamental Research, Mumbai 400005, India }
\affiliation {SUPA, University of the West of Scotland, Paisley, PA1 2BE, United Kingdom }
\affiliation {Institute of Astronomy, 65-265 Zielona G\'ora, Poland }
\affiliation {Indian Institute of Technology, Gandhinagar Ahmedabad Gujarat 382424, India }
\affiliation {Instituto de F\'\i sica Te\'orica, Univ. Estadual Paulista/ICTP South American Institute for Fundamental Research, S\~ao Paulo SP 01140-070, Brazil }
\affiliation {Andrews University, Berrien Springs, MI 49104, USA }
\affiliation {Trinity University, San Antonio, TX 78212, USA }
\affiliation {University of Washington, Seattle, WA 98195, USA }
\affiliation {Southeastern Louisiana University, Hammond, LA 70402, USA }
\affiliation {Abilene Christian University, Abilene, TX 79699, USA }
\fake{}

\begin{abstract}
We report results from a search for gravitational waves produced by perturbed intermediate mass black holes (IMBH) in data collected by LIGO and Virgo between 2005 and 2010. The search was sensitive to astrophysical sources that produced damped sinusoid gravitational wave signals, also known as ringdowns, with frequency $50\le f_{0}/\mathrm{Hz} \le 2000$ and decay timescale $0.0001\lesssim \tau/\mathrm{s} \lesssim 0.1$ characteristic of those produced in mergers of IMBH pairs. No significant gravitational wave candidate was detected. We report upper limits on the astrophysical coalescence rates of IMBHs with total binary mass $50 \le M/\mathrm{M}_\odot \le 450$ and component mass ratios of either 1:1 or 4:1. For systems with total mass $100 \le M/\mathrm{M}_\odot \le 150$, we report a 90\%-confidence upper limit on the rate of binary IMBH mergers with non-spinning and equal mass components of $6.9\times10^{-8}\,$Mpc$^{-3}$yr$^{-1}$. We also report a rate upper limit for ringdown waveforms from perturbed IMBHs, radiating 1\% of their mass as gravitational waves in the fundamental, $\ell=m=2$, oscillation mode, that is nearly three orders of magnitude more stringent than previous results.
\end{abstract}

\maketitle

\section{Introduction}\label{sec:overview}
Intermediate mass black hole (IMBH) binary systems represent a potential strong source of gravitational radiation accessible to ground-based interferometric detectors such as the Laser Interferometer Gravitational-Wave Observatory (LIGO)~\cite{2009ligopaper} and Virgo~\cite{2012virgopaper}. Although yet to be discovered, binary systems with total masses in the range $50\lesssim M/\mathrm{M}_\odot \lesssim10^5$ could form in dense star clusters such as globular clusters~\cite{2004millerimbhoverview, 2005imbhtimescale3, 2006gclifetime}.

The coalescence of a compact binary system generates a gravitational wave signal consisting of a low frequency inspiral phase when the compact objects are in orbit around each other, a merger phase marking the coalescence of the objects and the peak gravitational wave emission, and a high frequency ringdown phase after the objects have formed a single perturbed black hole~\cite{2006postnov, 2012faber}. For low mass systems, most of the signal-to-noise ratio comes from the inspiral phase of the coalescence. Several searches for gravitational waves from the inspiral of low mass compact objects have been performed by LIGO and Virgo~\cite{2009s5lowmass1yr, 2009s5lowmass186, 2012abadieS6lowmass}. However, since the merger frequency is inversely proportional to the mass of the system, it is shifted to lower frequencies for higher mass binaries. Searches for gravitational waves from the inspiral, merger and ringdown of binary black holes with total masses $25 \le M/\mathrm{M}_\odot \le100 $ have also been performed in LIGO-Virgo data~\cite{2011abadieS5highmass, 2013aasiS6highmass}.

For an IMBH binary, typically only the merger and ringdown parts of the signal fall above the low frequency cutoff of $40\,$Hz for the initial LIGO and Virgo detectors. Thus it is sufficient to conduct a search solely for these particular phases of the gravitational wave signal~\cite{1999creightonbhrd, 2005tamabhrd, 2005tsunesadabhrd}. A binary black hole merger is expected to result in a single perturbed black hole, and black hole perturbation theory and numerical simulations provide us with a well-understood ringdown signal model, a superposition of quasinormal modes that decay exponentially with time~\cite{1973teukolsky, 1970vish, 1970zerilli, 1971press, 1972price, 1976chandrasekar, 1984ferrari, 1999kokkotas}. Indeed, any perturbed black hole, not just that produced by a compact merger (e.g., a black hole formed as the result of the core collapse of a very massive star~\cite{2009popIIIstars, 2009VMStars, 2012VMStars2}), will emit ringdown gravitational waves described by its quasinormal modes. 

Since the gravitational waveform of perturbed black holes has a well-defined model, the method of matched filtering is used to search for ringdown signals. The first such search was carried out on data from the fourth LIGO science run (S4) which took place between February 22 and March 24, 2005~\cite{2009s4ringdown}. Additionally, two burst searches with less-constrained waveform models looked for gravitational waves from mergers of IMBHs in data collected by LIGO and Virgo between 2005 and 2010~\cite{2012s5imbh, 2013imbhs6}. No events were observed in these searches. In this paper, we present the results of a matched filter ringdown search of data from LIGO's fifth and sixth science runs and Virgo's science runs 2 and 3. We compare the resulting rate upper limits to the previous searches for gravitational waves from IMBHs.

Sections~\ref{sec:source} and~\ref{sec:waveform} describe the 
expected ringdown sources and waveform. Section~\ref{sec:data} provides
a brief description of the detectors and their sensitivities during the data collection epochs. Section \ref{sec:search} describes the search, and results are presented in Section 
\ref{sec:results}. Upper limits are presented in Section \ref{sec:ul} and discussed in Section \ref{sec:summary}.

\subsection{Ringdown sources}
\label{sec:source}
Observed black holes of known masses fall into two broad mass ranges. Stellar mass black holes have masses $\lesssim$~$35\,$M$_\odot$~\cite{2008ziolkowski, 2007prestwich, 2008silverman} although theoretical modeling of stellar evolution and population synthesis raises the possibility that significantly heavier stellar black holes could exist~\cite{belczynski10, 2012dominik}. Supermassive black holes have masses $\gtrsim$~$10^5\,$M$_\odot$~\cite{2003schodel, 2012bosch} and are thought to be cosmological in origin, possibly formed through galactic mergers leading to their growth through coalescences and accretion~\cite{2010volonteri, 2013supermassiveBH}. The large gap between the mass ranges of stellar and supermassive black holes is predicted to be populated by an elusive class of objects known as intermediate mass black holes (IMBHs)~\cite{1994IMBHfromgas, 2001IMBHfromPopIII, 2001imbhbinary2, 2002IMBHproduction,2004millerimbhoverview}. Observational evidence from ultra- or hyper-luminous X-ray sources and star cluster dynamics suggest a population of IMBHs with masses in the range $10^2\,$M$_\odot$ to $10^4\,$M$_\odot$~\cite{2004millerimbhoverview}. Ultra-luminous X-ray sources with angle-averaged fluxes many times that of a stellar mass black hole accreting at the Eddington limit $(> 3\times10^{39}\,$erg$\,$s$^{-1})$ may be explained by black holes with masses larger than any known stellar mass black hole. The brightest known  hyper-luminous X-ray source and the strongest IMBH candidate is the point-like X-ray source HLX-1. Its maximum X-ray luminosity of $10^{42}\,$erg$\,$s$^{-1}$ requires a black hole mass $\gtrsim$~a~few $10^3\,$M$_\odot$~\cite{2009imbhhlx1, 2012imbhhlx1}. Other hyper-luminous X-ray sources include M82 X-1~\cite{2001m82}, Cartwheel N10~\cite{2010cartwheel}, and CXO J122518.6~\cite{2010cxo}. Furthermore, the excess of dark mass at the centers of globular clusters could be explained by $\sim10^3\,$M$_\odot$ IMBHs formed from repeated mergers between other compact objects and/or stars~\cite{2002gcdynamics1, 2002gcdynamics2, 2005gcdynamics3}. However, both hyper-luminous X-ray sources and central globular cluster masses can be explained via phenomena that do not include IMBHs~\cite{2003noimbh, 2005noimbhulx}. Still, most observational evidence for globular cluster IMBHs using radio emissions can place upper bounds of $\le10^3\,$M$_\odot$~\cite{bozzo11,dalessandro11,kirsten12,mcnamara12,strader12,wrobel11}, and do not rule out lower mass systems that are above the expected maximum mass of a normal stellar mass black hole~\cite{belczynski10}. Thus, the existence of IMBHs currently remains speculative.

Numerical simulations suggest that IMBH binaries could form in collisional runaway scenarios in young dense star clusters. Initially, in young star clusters, IMBHs could form via the runaway collapse of very massive stars~\cite{1999imbhbinary1, 2001imbhbinary2, 2000imbhbinary3, 2004imbhbinary4}. After separate formation, two IMBHs could settle to the core of the cluster through dynamical friction and form a common binary via dynamical interactions. The binary would tighten due to three-body encounters, finally merging quickly via gravitational radiation~\cite{1996imbhtimescale1, 2003imbhtimescale2, 2005imbhtimescale3}.

From~\cite{2010rates}, we know that the astrophysical rate of IMBH binary coalescence in globular clusters (GC) should be no higher than $0.07\,$GC$^{-1}$Gyr$^{-1}$ assuming that all globular clusters are sufficiently massive and have a sufficient binary fraction to form this type of binary once in their lifetime of $13.8\,$Gyr~\cite{2006gclifetime}. Also, globular clusters have a space density of roughly $3\,$GC$\,$Mpc$^{-3}$~\cite{2008gcdensity}. This allows us to convert the astrophysical upper limit to $2\times10^{-10}\,$Mpc$^{-3}$yr$^{-1}$. If we assume that only 10\% of globular clusters meet these requirements, the rate would still be as high as one tenth this value~\cite{2010rates}. 

Numerical simulations also suggest the possibility of forming intermediate mass ratio inspirals (IMRIs) (e.g., a coalescence of an IMBH with a compact stellar mass companion) in these same dense star clusters. This occurs through a combination of gravitational wave emission, binary exchange processes, and secular evolution of hierarchical triple systems~\cite{2000taniguchi, 2002mouri, 2002IMBHproduction, 2004gultekin, 2007oleary}. Ringdown searches in the advanced detector era could be important for detecting IMRIs, particularly if the inspiraling companion is a black hole with $m \gtrsim$~$10\,$M$_\odot$ or if the system is a compact object coalescing with a slowly-spinning IMBH with $m \gtrsim$~$350\,$M$_\odot$~\cite{2008gcdensity}.

\subsection{Ringdown waveform}
\label{sec:waveform}
A black hole can be perturbed in a variety of ways, e.g., by interaction with a companion, by accretion or infall of matter, or in its formation through asymmetric gravitational collapse. A perturbed Kerr black hole will emit gravitational waves, relaxing to a stable configuration through radiation generated by a superposition of  quasinormal modes of oscillation~\cite{1973teukolsky, 1970vish, 1970zerilli, 1971press, 1972price, 1976chandrasekar, 1984ferrari, 1999kokkotas}. The emitted gravitational waves are exponentially decaying sinusoid signals characterized  by a complex angular frequency $\omega_{\ell mn}$ from which we can derive both the real frequency $f_{\ell mn}$ and the quality factor $Q_{\ell mn}$:
\begin{eqnarray}\label{eq:qflm}
f_{\ell mn}=\Re(\omega_{\ell mn})/2\pi\, ,\\
Q_{\ell mn}=\pi f_{\ell mn}/\Im(\omega_{\ell mn})\, ,
\end{eqnarray}
 where $\ell=2,3,...,$ and $m=-\ell,...,\ell$ are the spheroidal harmonic indices and $n$ denotes the overtones of each mode. Overtones with $n>0$ are generally negligible in amplitude compared with the fundamental $n=0$ mode. Numerical simulations have demonstrated that the $\ell=m= 2$ fundamental mode dominates the gravitational wave emission, particularly in the case of an equal mass compact object merger~\cite{2007domlmn}. The ringdown search uses single-mode waveform templates. However, other modes can contribute significantly to the gravitational wave signal, particularly in cases where the binary's mass ratio $q=m_>/m_<\neq1$ where $m_>=\mathrm{max}(m_1, m_2)$ and $m_<=\mathrm{min}(m_1, m_2)$. Reference~\cite{2007rdparams} reports that single-mode templates can result in a loss $\gtrsim 10\%$ in detected events over a significant mass range and also result in large errors in the estimated values of parameters (especially the quality factor). A multimode ringdown search would perform better both in efficiency and parameter estimation~\cite{2007rdparams}. Nevertheless, we show that the single-mode ringdown search will still provide good sensitivity to comparable mass binary systems (see average sensitive distances given in Section~\ref{sec:eobul}).
 
The response of an interferometric detector to a gravitational wave is
\begin{equation}\label{eq:hoft}
h(t)= F_+(\theta, \phi, \psi)h_+(t) + F_\times(\theta, \phi, \psi) h_\times(t)
\end{equation}
where $F_+$ and $F_\times$ are the antenna pattern functions that depend on the direction to the source as described by a polar angle $\theta$, an azimuthal angle $\phi$, and a polarization angle $\psi$. The plus and cross polarizations $h_+$ and $h_\times$ of a single-mode $(\ell, m, n) = (2,2,0)$ ringdown waveform take the approximate form
\begin{equation}\label{eq:hplus}
\begin{split}
 h_+(t; \iota, \phi) = &\frac{\mathcal{A}}{r} \left( 1 + \cos^2\iota \right) e^{-\pi f_0 \left(t-t_0\right) / Q} \\
 & \cos\left[ 2\pi f_0 \left(t - t_0 \right) + \phi_0 \right],
 \end{split}
 \end{equation}
 \begin{equation}\label{eq:hcross}
 \begin{split}
 h_\times(t; \iota, \phi)  =  &\frac{\mathcal{A}}{r} \left( 2\cos \iota \right) e^{-\pi f_0 \left(t -t_0\right) / Q} \\
 &\sin\left[ 2\pi f_0 \left(t - t_0 \right) + \phi_0 \right],
 \end{split}
\end{equation}
for $t>t_0$ where $f_{0}=f_{220}$ and $Q=Q_{220}$ are the oscillation frequency and the quality factor of the $(\ell, m, n) = (2,2,0)$ mode,  $r$ is the distance to the source, $\phi_0$ is the initial phase of the mode, and $\iota$ is the inclination angle. The oscillation amplitude of the $(\ell, m, n) = (2,2,0)$ mode, ${\cal A}$, is given approximately by (see Appendix~\ref{sec:appendix1})
\begin{equation}\label{eq:amplitude}
{\cal A}=\frac{GM}{c^2} \sqrt{\frac{5\epsilon}{2}}\,\,Q^{-1/2}F(Q)^{-1/2}g(\hat{a})^{-1/2}\,,
\end{equation} 
where $G$ is the gravitational constant, $M$ is the black hole mass, $c$ is the speed of light, $\epsilon$, known as the ringdown efficiency, is the fraction of the black hole's mass radiated, $\hat{a}=cS/GM^2$ where $S$ is the black hole's spin angular momentum, $F(Q)=1+1/(4Q^{2})$ and $g(\hat{a})=\left[1.5251-1.1568(1-\hat{a})^{0.1292}\right]$ [cf. Eq.~(\ref{eq:f0}),~(\ref{eq:q}), and~(\ref{eq:rdamp})].

The total ringdown efficiency of a black hole binary with non-spinning components is known to scale with the square of the symmetric mass ratio, $\nu=m_1m_2/(m_1+m_2)^2=q/(1+q)^2$, as
$\epsilon\approx0.44\nu^2$~\cite{2007etaeqn, 2012kamaretsosprl, 2012kamaretsosprd}. Thus, for $q=1$, $\epsilon\sim3\%$ and, for $q=4$, $\epsilon\sim1\%$. Gravitational waves from extreme mass ratio systems will not be detectable unless the system is sufficiently close (see Section~\ref{sec:eobul}). A black hole binary with spinning components will radiate more energy if the spins are aligned with the orbital angular momentum and less if the spins are anti-aligned~\cite{2006orbhangup, 2012kamaretsosprl}.

The black hole mass $M$ and dimensionless spin parameter $\hat{a}$ can be determined numerically using fitting formulae to Kerr quasinormal mode frequency and quality factor parameters tabulated in Table VIII of~\cite{2006bertiqnm}. For the $(\ell, m, n) = (2,2,0)$ mode, the fits are of the form:
\begin{eqnarray}
f_0 &=& \frac{1}{2\pi}\frac{c^3}{GM}\left[1.5251-1.1568\left(1-\hat{a}\right)^{0.1292} \right]\, , \label{eq:f0}\\
Q &=& 0.7000+1.4187\left(1-\hat{a}\right)^{-0.4990}\, . \label{eq:q}
\end{eqnarray}
These fitting functions allow us to relate a measurement of the frequency and quality factor from a match filter ringdown template to the mass and angular momentum of the final perturbed black hole.

We can approximate the ringdown gravitational wave strain by
\begin{equation}\label{eq:waveform}
h_{0}(t)={\cal A_\mathrm{eff}}\,e^{-\pi f_{0}(t-t_0)/Q}\cos[2\pi f_{0}(t-t_0) + \varphi_{0}]\, ,
\end{equation}
for $t>t_0$ where $\cal A_\mathrm{eff}=\cal A/D_\mathrm{eff}$ and  $D_\mathrm{eff}$ is the effective distance to the source and $\varphi_{0}$ is the effective initial phase depending on the initial phase $\phi_0$ as well as on the signal polarization [see Eq.~(1.7) and~(1.9) in~\cite{Talukder:2013ioa}]. Note that both $\varphi_{0}$ and time of arrival at the detector $t_0$ are set to zero for simplicity in the template waveform given in Section~\ref{sec:algorithm}.

\section{Data Set}
\label{sec:data}
The data analyzed spans multiple science runs for both the LIGO and Virgo detectors. We report results both for data collected between November 2005 and September 2007 and between July 2009 and October 2010.

The first time period covers LIGO's fifth science run (S5). The LIGO site in Hanford, Washington hosted two collocated interferometers: a $4\,$km detector H1 and a $2\,$km detector H2. The LIGO site in Livingston, LA hosted one $4\,$km detector L1. Additionally, the Virgo $3\,$km detector in Cascina, Italy operated from May 2007 to September 2007 during its first science run (VSR1) which overlapped with the last few months of LIGO's S5 run. However, this search did not analyze VSR1 data. Thus, for the first time period, which we designate Period~1, we report results for the three-fold coincident search of the H1H2L1 detector network. We also report results for two-detector combinations of this network including H1L1 and H2L1. We chose to exclude H1H2 coincident events since accurately measuring the significance of gravitational wave candidates is complicated by this network's correlated detector noise. 

The second time period covers LIGO's sixth science run (S6) during which only the H1 and L1 LIGO detectors were operating. The Virgo detector conducted two science runs during this period: VSR2 which ran from July 2009 to January 2010, and VSR3 which ran from August 2010 to October 2010. For this second time period, which we designate Period~2, we report results for the coincident search of the H1L1V1 detector network. We also report results for all two-detector combinations within this network.

\begin{figure*}
\centerline{\includegraphics[width=0.5\textwidth]{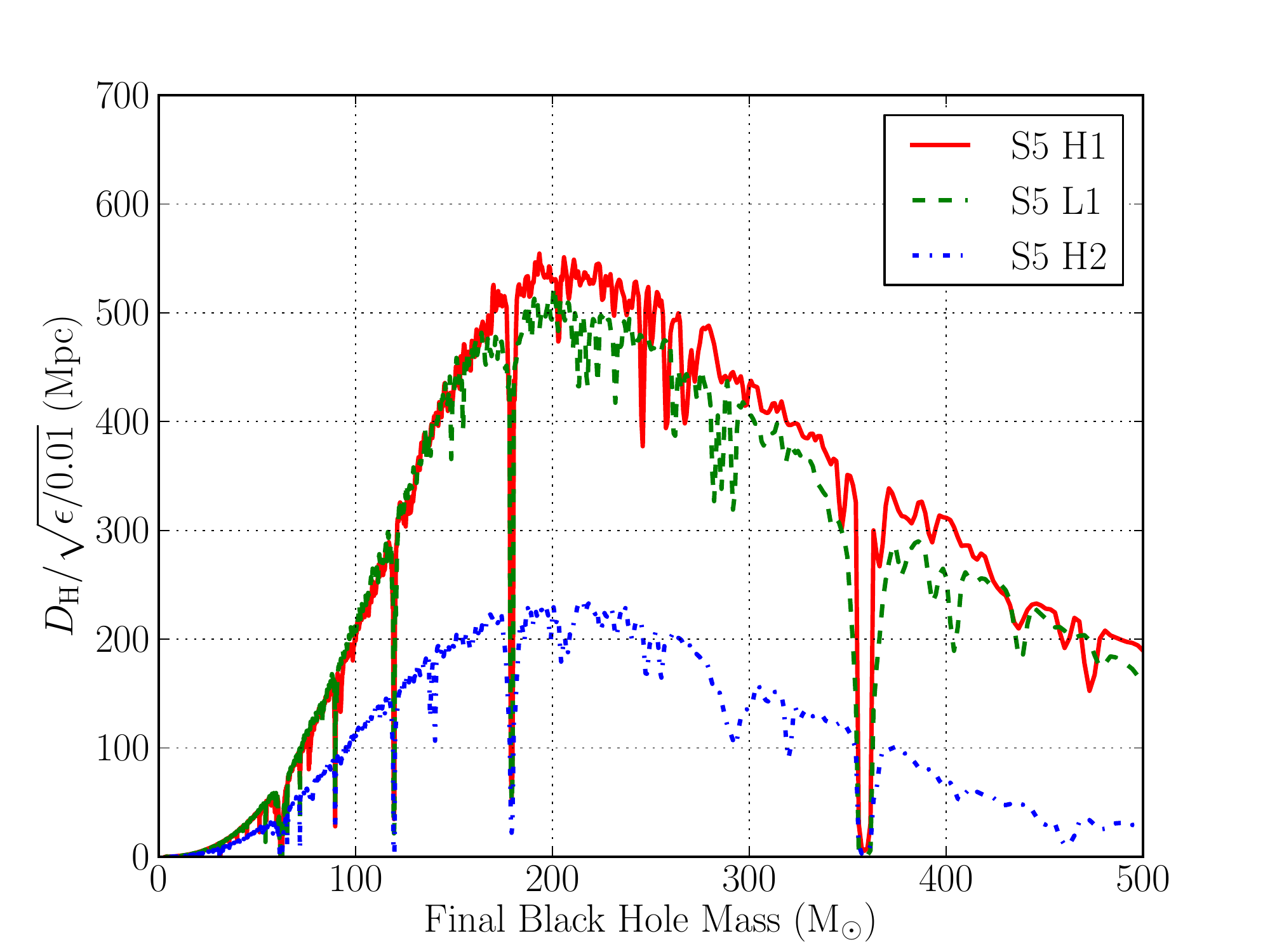}
  \includegraphics[width=0.5\textwidth]{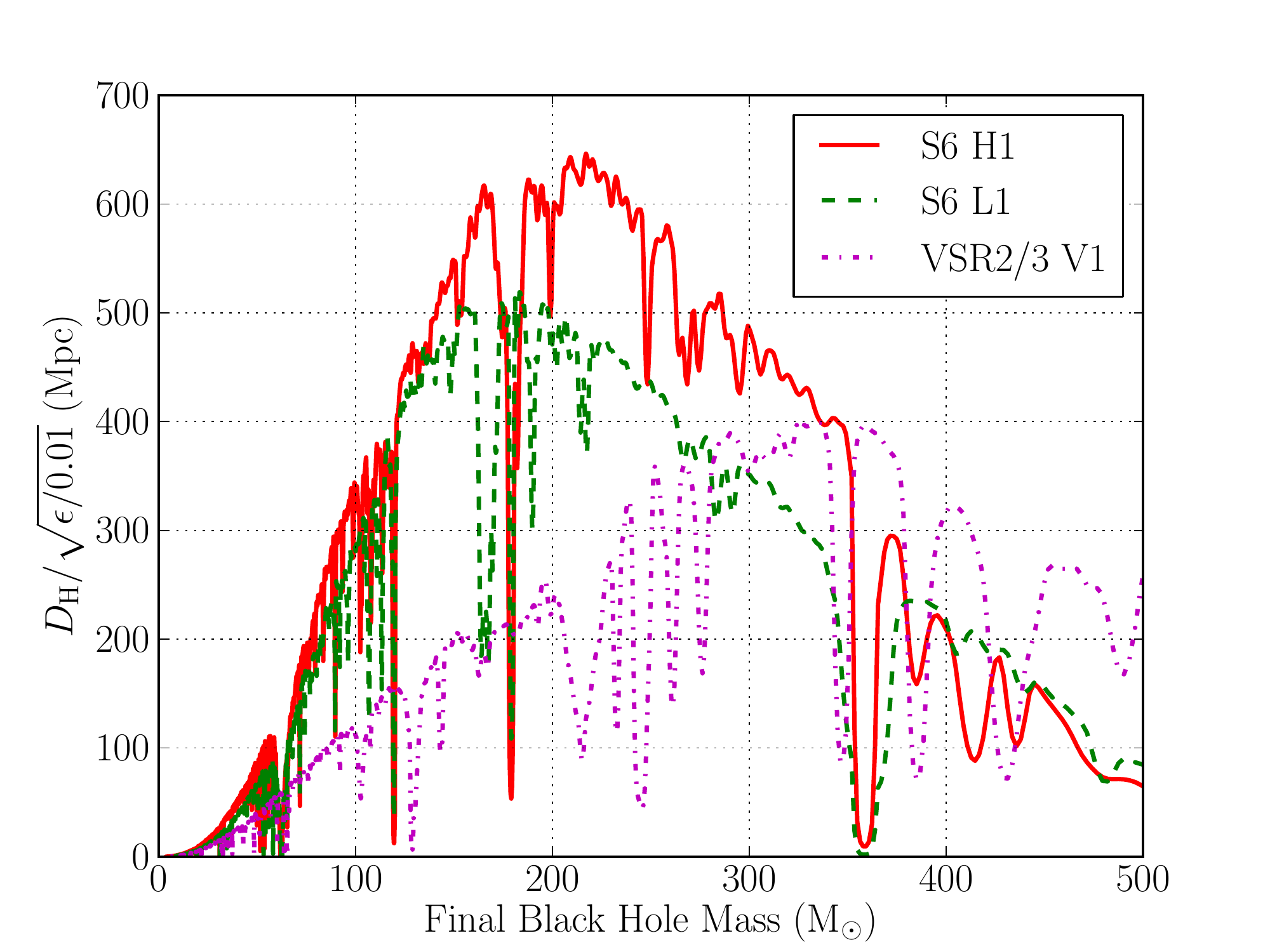}
  \label{fig:rdhorizon}}
  \caption{Ringdown horizon distances, $D_\mathrm{H}$, weighted by the square root of the ringdown efficiency, $\sqrt{\epsilon}$, as a function of final black hole mass for Period~1 ({\it left}) and Period~2 ({\it right}). Here we have set $\epsilon=1\%$. The dimensionless spin parameter is set to $\hat{a}=0.9$. For example, during Period~1, a $\sim200\,$M$_\odot$ ringdown source with $\epsilon=1\%$, $\hat{a}=0.9$, and optimal location and orientation at a distance of $\sim530\,$Mpc would produce a signal-to-noise ratio of 8 in the H1 detector.}
  \label{fig:sensitivity}
\end{figure*}

LIGO's S5 run marked the final data collection of the {\it initial} LIGO detector configuration during which design sensitivity was achieved~\cite{2009ligopaper}. Figure~\ref{fig:sensitivity} ({\it left}) demonstrates the H1, H2, and L1 detectors' sensitivities to ringdown signals from spinning black holes with $\hat{a}=0.9$ and $\epsilon=1\%$\footnote{These values were chosen so that a direct comparison could be made with Fig.~2 in~\cite{2009s4ringdown}.} for typical Period~1 performance. This figure shows the horizon distance $D_\mathrm{H}$ divided by the square root of the ringdown efficiency $\epsilon$, scaled to a canonical value $\epsilon=1\%$, as a function of the final black hole mass.  The horizon distance is the distance at which a given source with optimal location and orientation would produce a SNR of 8 in a given detector; some details of its derivation for ringdowns are given in Appendix~\ref{sec:appendix2}. Dips in the ringdown horizon distance correspond directly to features of the detectors' noise spectral density curves. For instance, the strong dip in sensitivity at $360\,$M$_\odot$ is due to $60\,$Hz electric power noise.

The S6 run, during the phase of the {\it enhanced} LIGO detector configuration, followed a series of upgrades to the initial detectors to improve sensitivity. These enhancements included a higher power laser and a new DC readout system~\cite{2009eLIGO}. Similarly, the Virgo detector saw several improvements between its VSR1 and VSR2 runs including a more powerful laser, a thermal compensation system, and improved scattered light mitigation.  Before Virgo's VSR3 run in early 2010, monolithic suspensions with fused-silica fibers were installed~\cite{2011virgoVSR2}. Figure~\ref{fig:sensitivity} ({\it right}) demonstrates the H1, L1, and V1 detectors' sensitivities to ringdown signals from spinning black holes with $\hat{a}=0.9$ and $\epsilon=1\%$ for typical Period~2 performance.

Gravitational-wave strain data from each of the detectors are known to be both non-Gaussian and non-stationary. Non-Gaussianity is often manifested as noise transients, or glitches, in the strain data. Efforts are made to diagnose and remove glitches and stretches of elevated noise from the data set using environmental and instrumental monitors~\cite{2008glitchgroup, 2010s6dq, 2010upveto}. In this search, as in previous searches of LIGO-Virgo data, we apply three levels of data quality vetoes~\cite{2010jakethesis, 2010catveto} (see Appendix~A of~\cite{2009s5lowmass1yr} for more details). Data remaining after the first and second veto levels have been applied are searched for possible detection candidates (see Section~\ref{sec:results}). Data remaining after all three veto levels have been applied are searched for detection candidates and are also used in constraining the IMBH merger rate (see Section~\ref{sec:ul}).  Table~\ref{tab:livetime} gives the total analyzed time after all three veto levels are applied and after the removal of the ``playground" data set used for pipeline tuning as described in Section~\ref{sec:sim}. The total analysis time for both Period~1 and Period~2 was  $1.2\,$years.

\begin{table}[t]
\caption{Length of each network's total analyzed time after the third level
of vetoes has been applied and the playground data set has been removed.}
\label{tab:livetime}
\begin{ruledtabular}
\begin{tabular}{lcc}
 & \multicolumn{2}{c}{Analysis Time\footnote{Excluding playground time.} (years)} \\
Network	& Period 1 & Period 2 \\
\colrule
H1L1	&	0.09	&	0.17	\\
H1V1	&	 --	&	0.10	\\
H2L1	&	0.07	&	 --	\\
L1V1	&	 --	&	0.06	\\
H1H2L1	&	0.63	&	 --	\\
H1L1V1	&	 --	&	0.08	\\
\hline
Total          &      0.79  &      0.41 \\
\end{tabular}
\end{ruledtabular}
\end{table}

\section{Ringdown Search}
\label{sec:search}

\subsection{Search Algorithm}
\label{sec:algorithm}
The ringdown search algorithm, first introduced in~\cite{1999creightonbhrd, 2009s4ringdown}, is based on the optimal method for finding modeled signals buried in Gaussian noise, the matched filter~\cite{helstrom1960statistical}. The data from multiple gravitational wave detectors are match filtered with single-mode ringdown templates to test for the presence or absence of signals in the data. The output is a signal-to-noise ratio (SNR) time series~\cite{2009s4ringdown} from which local maxima above a pre-determined SNR threshold, called triggers, are retained for further analysis. Since the noise in the detector data is non-stationary and non-Gaussian, matched filtering alone is not enough to establish that a trigger is a gravitational wave signal. Since detector noise can often mimic the signal for which we are searching, additional tests are employed including detector coincidence and SNR consistency. We use a search pipeline similar to the {\tt ihope} pipeline described in~\cite{2013ihope}. Here we summarize the main steps of the ringdown search pipeline.

The data conditioning and segmentation is discussed in detail in~\cite{2008gogginthesis}.  Each segment of data is filtered using a bank of ringdown templates characterized by frequency $f_0$ and quality factor $Q$. Following~\cite{2009s4ringdown}, the template used in this search is
\begin{equation}\label{eq:template}
h(t) = e^{-\frac{\pi f_{0}t}{Q}}\cos(2\pi f_{0}t)\,, \;\;\;\;\;\;0\leq t\leq t_{\mathrm{max}}
\end{equation}
[cf. Eq.~(\ref{eq:waveform})], with a length of 10 e-folding times, $t_{\mathrm{max}}=10Q/\pi f_{0}$.\footnote{An arbitrary initial phase parameter (or equivalently, a quadratic sum of sine and cosine template outputs) could be implemented in the template waveform to reduce the fraction of power lost in the event of a pure sine wave signal. The problem is most acute for the detection of perturbed black holes with high frequency ($f_0\gtrsim 1000\,$Hz) and low dimensionless spin parameter ($\hat{a}\lesssim 0.6$) where significant power is lost by using a cosine template~\cite{Talukder:2013ioa}. However, allowing an arbitrary phase would increase the noise level of the search. Furthermore, any ringdown signal would follow a preceding waveform and there is some arbitrariness in the division of one from the other.}

The template bank is tiled in ($f_0$, $Q$)-space according to the analytic approximate metric computed assuming white detector noise as described in~\cite{2003tamard, 2005tamabhrd, 2009s4ringdown} so that no point in the parameter space has an overlap of less than 97\% with the nearest template.\footnote{The template placement metric is derived using a sine template in~\cite{2003tamard} whereas a cosine template is used to filter the data. Optimally, the metric derivation should account for initial phase dependence as derived in~\cite{2004improverdtmplt}. In the high~{\it Q} limit, the sine and cosine metrics coincide.} The template parameters cover a frequency band between $50\,$Hz and $2\,$kHz and quality factor in the physical range between 2 and 20. This corresponds roughly to masses in the range $10\,\mathrm{M}_\odot$ to $600\,\mathrm{M}_\odot$, and spins in the range 0 to 0.99. A fixed bank of 616 templates was used for all detectors. 

Triggers with an SNR statistic above a predetermined threshold $\rho^*$ are retained for further analysis. For both Period~1 and Period~2, we set $\rho^*_\mathrm{H1}=\rho^*_\mathrm{L1}=5.5$. For the least sensitive detector in each analysis period, we set lower thresholds: $\rho^*_\mathrm{H2}=4.0$ and $\rho^*_\mathrm{V1}=5.0$.

\subsection{Coincidence and Vetoes}
\label{sec:coincidence}
Once triggers are found in a single detector, we apply a coincidence test, analogous to the one introduced in~\cite{ethinca2008}, to check for multi-detector parameter and arrival time consistency. In order to include information about time coincidence d$t$ and template coincidence for d$f_0$ and d$Q$ in a single coincidence test, we construct a 3D-metric~\cite{2003tamard} to calculate the distances d$s^2$ between two triggers in ($f_{0}$, $Q$, $t$)-space. The quantity $(1-\mathrm{d}s^2)$ is a measure of normalized signal mismatch. To account for the finite travel time between non-collocated detectors, we minimize d$s^2$ for each detector pair over a range of allowed time differences. Only pairs of triggers for which  d$s^2 \le \mathrm{d}s_*^2=0.4$ are kept as coincident candidates. During times when three detectors are operating, triple coincident events are constructed from sets of three triggers if each trigger in the set passes the coincidence test with every other one. We also consider H1L1 coincidences in a H1H2L1 network.

We also apply second and third level vetoes to segments of poor data quality as described in~\cite{2013ihope}. Additionally, for Period~1, we apply a number of amplitude consistency tests that exploit the coalignment of H1 and H2~\cite{2013ihope}. These tests allow us to apply cuts to reduce the background of false alarms.

\subsection{Ranking Events}
\label{sec:ranking}
Finally, the pipeline ranks the coincidences and determines significance. For this purpose, a detection statistic is designed to separate signal-like coincidences from noise-like coincidences. Given the large number of parameters that describe multi-detector coincidences, we employ a multivariate analysis using cuts on multiple parameters to help in classifying coincidences as signals or false alarms: i.e.,~a multivariate statistical classifier. The parameters provided to the classifier to aid in characterizing the multi-detector coincidences included single-detector SNRs and differences in time and template parameters between detectors, recovered effective distances, composite SNR statistics,\footnote{Some details of the composite SNR statistics used for classification are given in~\cite{Talukder:2013ioa}.} the 3D-metric distance between triggers and the metric coefficients as well as data quality information from the hierarchical veto method described in~\cite{2011hveto}. Additional details of these parameters will be described in a future paper.

To perform the multivariate analysis, we use a machine learning algorithm known as a random forest of bagged decision trees~\cite{2001breiman, 2007Narsky}. Similar techniques have been implemented for detecting gravitational-wave bursts~\cite{2013burstmvsc} and cosmic strings~\cite{2013cosmicstring}. The training of the classifier uses two sets of data: a collection of coincidences associated with simulated signals and a collection of accidental coincidences that act as a proxy for the background.

The simulated signal set is generated by adding software-generated gravitational waveforms to the data and running a separate search. The simulated waveforms, described in more detail in Section~\ref{sec:sim}, included both full coalescence IMBH merger signals and lone ringdown signals.

The set of accidental coincidences is generated using the method of time-shifted data that takes advantage of the fact that a real signal will produce triggers in each detector that are coincident in time. The data streams of detectors are shifted in time with respect to one another by intervals longer than the light travel time between sites plus timing uncertainties, then a search for coincidences is performed. These time-shifted coincidences are then almost certainly due to noise. For Period~1, the L1 data stream was shifted by multiples of 5 seconds relative to H1 and H2 for a total of 100 time-shifted analyses; the H1 and H2 data streams were not time-shifted relative to one another. For Period~2, the L1 data stream was shifted by multiples of 5 seconds and the V1 data stream was shifted by multiples of 10 seconds relative to H1 for a total of 100 time-shifted analyses.

The classifier assigns a likelihood ranking statistic $\mathcal{L}$ to each coincidence. A high likelihood implies the coincidence is signal-like; a low likelihood implies the coincidence is noise-like. For each candidate, we need to be able to assign a significance to its likelihood ranking. This is done by mapping a false alarm rate (FAR) to a candidate's rank in order to assess its significance. We count the number of false coincidences in the time-shifted searches, record their likelihood values, and determine the analysis time $T_b$ of all the time-shift searches for a particular experiment time (e.g., H1L1 coincidences in a H1L1V1 network, H1L1V1 coincidences in a H1L1V1 network, etc.). We perform this calculation separately for each type of coincidence in each of the different experiment times. Then, for each candidate in each experiment, we determine the FAR at its likelihood value $\mathcal{L}^*$ with the expression:
\begin{equation}\label{eq:far}
\mathrm{FAR}=\frac{\sum\limits_{k=1}^{100} N_k(\mathcal{L} \ge \mathcal{L}^*)}{T_b}
\end{equation}
where $N_k$ is the measured number of coincidences with $\mathcal{L} \ge \mathcal{L}^*$ in the $k^{th}$ shifted analysis. We performed a total of 100 time-shifted analyses. Finally, we can rank candidates by their FARs across all types of experiment times into a combined ranking, known as combined FAR, for a single experiment time as described in detail in~\cite{2009keppelthesis}. The combined FAR is the final detection statistic that allows us to combine the candidate rankings from the various experiment types into a single list of candidates ordered from most significant to least significant.

\subsection{Tuning and simulations}
\label{sec:sim}
The analysis was tuned using the set of false alarm coincidences obtained from time-shifted searches, a set of simulated signals (``injections") added to the detectors' data streams in a separate stage of data analysis, and a small chunk of the actual search data, approximately 10\%, designated ``playground", that was later excluded from the analysis to preserve blindness. The goal of tuning the analysis is to maximize the sensitivity of the search while minimizing the false alarm rate. For this, we injected a set of ringdown-only waveforms with $\epsilon=1\%$ into the data set. The waveforms were determined by Eq.~(\ref{eq:hoft}),~(\ref{eq:hplus}), and~(\ref{eq:hcross}) with sky location and source orientation sampled from an isotropic distribution. Several sets of ringdown waveforms were injected with a uniform distribution in $f_0$ and $Q$ to cover the parameter range of the ringdown template bank. Also,  in order to cover the broad mass and spin range accessible to the ringdown search when signals have $\epsilon=1\%$, several sets of ringdown waveforms were injected with a uniform distribution in $M$ and $\hat{a}$: $50 \le M/\mathrm{M}_\odot \le 900$ and $0.0 \le \hat{a} \le 0.99$. Additionally, we also injected a set of full coalescence waveforms with isotropically-distributed sky location and source orientation parameters into the data. These full coalescence waveforms included the recently-implemented non-spinning EOBNRv2 family~\cite{2011eobnrv2} and the spinning PhenomB family~\cite{2011IMRPhenomB}. The EOBNRv2 injections were distributed uniformly in total mass $50 \le M/\mathrm{M}_\odot \le 450$ and in mass ratio $1 \le q \le 10$. The PhenomB injections were given the same mass distribution and a uniform dimensionless spin parameter $0.0 \le \hat{a}_{1,2} \le 0.85$ where $\hat{a}_{1,2}=cS_{1,2}/Gm_{1,2}^2$ for the spin angular momentum $S$ and the mass $m$ of the two binary components. For a discussion of the injection sets used in computing rate upper limits, see Section~\ref{sec:ul}.

\section{Search results}
\label{sec:results}
\begin{figure}
\centerline{\includegraphics[width=0.5\textwidth]{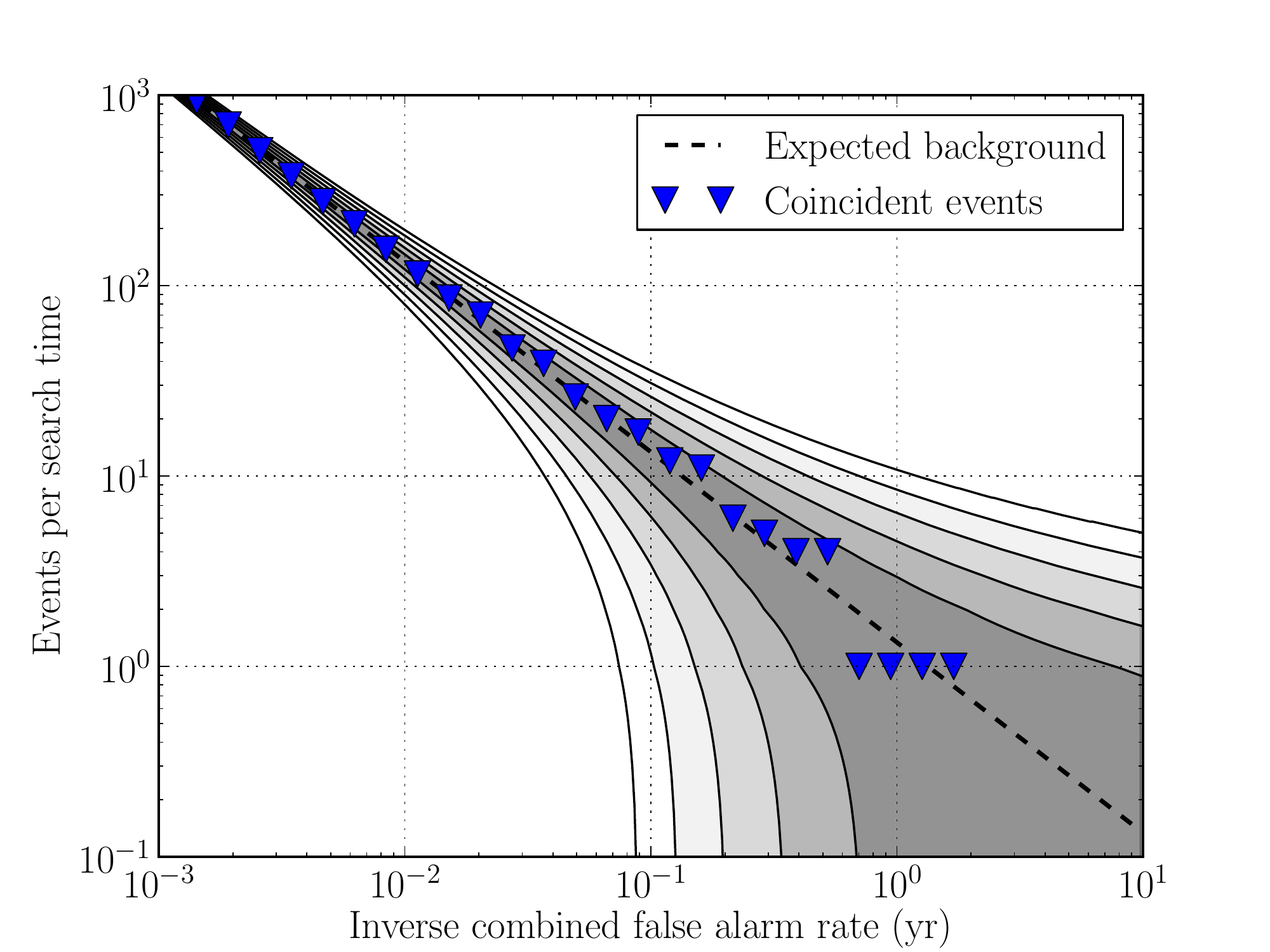}}
  \caption{Cumulative distributions of coincident events found as a function of inverse combined false alarm rate after all vetoes up to the third level are applied. The figures combines results from both triple and double coincident searches over the total analysis time of Period~1 and Period~2. Grey contours mark the 1$\sigma$ through 5$\sigma$ region of the expected background from accidental coincidences. No search candidates stand out from the background.} \label{fig:IFAR}
\end{figure}

The search yielded no significant gravitational wave candidates,  as all events were consistent, within 1 sigma, with the background from accidental coincidences. Figure~\ref{fig:IFAR} shows the cumulative distributions of coincident events found as a function of inverse combined false alarm rate after all vetoes up to the third level are applied. These plots combine results from both triple and double coincident searches over the total analysis time of Period~1 and Period~2. 

The most significant event was found in triple coincidence during Period~1 in H1, H2, and L1. After the first and second level vetoes were applied, it was found with a combined $\mathrm{FAR}=2.07\,$yr$^{-1}$ and, after the third level vetoes were additionally applied, with a combined $\mathrm{FAR}=0.45\,$yr$^{-1}$. Thus we expect an accidental coincidence to be found by the search with this significance $\sim$ once per two years of analysis. Since the total analysis time was $1.2\,$years, the event is consistent, within 1 sigma, with the accidental coincidence rate. In both H1 and H2, a trigger was found barely above threshold with matched filter SNRs of 5.5 and 4.4, respectively. However, the candidate was found as a very loud trigger in L1 with a matched filter SNR of 48.9. Performing a coherent Bayesian parameter estimation follow-up~\cite{S6parameterestimation} on these triggers, we found that a coherent analysis favored a solution for the binary's sky location and orientation that yield a very strong signal in L1, but virtually no response in H1 and H2 detectors.  While it is theoretically possible that very particular location and orientation parameters could produce such a signal, an excursion from stationary, Gaussian noise (a glitch) in L1 is more likely.

\section{Rate Limits}
\label{sec:ul}
In this section, we compute the 90\%-confidence upper limits on IMBH coalescence rates and IMBH black hole ringdown rates. The former will allow us to make an astrophysical statement as well as to compare the sensitivity of the ringdown search to various other searches that have made statements in this mass regime, including~\cite{2011abadieS5highmass,2012s5imbh, 2013aasiS6highmass, 2013imbhs6}.

We used a procedure similar to that discussed in~\cite{2011abadieS5highmass,2013aasiS6highmass} for the upper limit calculation based on the loudest event statistic~\cite{2009biswasloudest, 2013corrigendum}. In order to capture the variability of the detector noise and sensitivity, we analyzed the data in periods of $\sim1$ to $2$ months. In each of these analysis times, we estimate the volume to which the ringdown search is sensitive by injecting many simulated signals into the data and performing an analysis to recover them. In Section~\ref{sec:eobul}, we describe the distribution of EOBNRv2 waveforms used to model the source population of IMBH binaries. Our sensitivity to these signals depends on total mass, mass ratio, source distance, and sky location as well as other parameters such as component spins. We explore the changing sensitivity of the ringdown search to these binaries over a range of total masses for both equal mass and 4:1 mass ratio systems. Other distance and orientation parameters are randomly sampled. Due to the significant variation of the search sensitivity over the large mass and mass ratio parameter space that we explore in Section~\ref{sec:eobul}, we have chosen to include only systems with non-spinning components in this study. In Section~\ref{sec:rdul}, we describe the distribution of ringdown waveforms used to model the population of perturbed black holes first explored in~\cite{2009s4ringdown}.

For each of these injection sets, we compute the sensitive volume for a given mass range and mass ratio by integrating the efficiency of the search over distance:
\begin{equation}\label{eq:veff}
V_\mathrm{eff}=4\pi\int \eta(r) r^2 dr
\end{equation}
where the efficiency $\eta(r)$ is calculated as the number of injections found with a lower combined FAR than the most significant coincident event in each analysis time for the search divided by the total number of injections made at a given distance. As described in~\cite{2009biswasloudest,2011abadieS5highmass,2013aasiS6highmass, 2013corrigendum}, we estimate the likelihood parameter $\Lambda$ of the loudest event being a signal versus being caused by an accidental coincidence for each type of coincident network time and each mass and mass ratio bin. For each analysis time (excluding playground time), effective volume from Eq.~(\ref{eq:veff}), and estimated $\Lambda$, we marginalize over statistical uncertainties given in Section~\ref{sec:uncertainty} and construct a marginalized likelihood as a function of the astrophysical rate in units of mergers per Mpc$^3$ per year for our EOBNRv2 injection sets and in units of ringdowns per Mpc$^3$ per year for our ringdown injections. In order to obtain a combined posterior probability distribution for the rate over all the analysis times, we multiply a prior on the rate by the product of the marginalized likelihood functions to obtain a posterior probability and integrate to 90\% to obtain the 90\%-confidence upper limit on the rates. For our combined Period~1 result, we assumed a uniform prior on the rate. However, for the main Period~2 result, we were able to use the Period~1 posteriors over coalescence or ringdown rate as priors for the upper limit calculation.

\subsection{Sources of uncertainty}
\label{sec:uncertainty}
We must account for several sources of random and systematic error when computing rate upper limits. Uncertainties on the sensitive volume as well as incomplete knowledge of waveforms and source populations form the largest contributors. As described in earlier search papers~\cite{2012abadieS6lowmass,2011abadieS5highmass,2013aasiS6highmass}, we marginalize over random uncertainty (i.e. calibration and statistical Monte Carlo uncertainties) for each analysis time. The 90\%-confidence upper limits based on the marginalized posterior distributions are the main results of this search.

The calibration of the data is a source of both random and systematic error. Reference~\cite{2010abadieS5calib} reports uncertainties on the magnitude of the response function for each detector in Period~1. We find an overall distance uncertainty of 8\%. Thus, the random uncertainty on the visible volume for Period~1 is approximately 8\% cubed, or 24\%. For Period~2, references~\cite{2011bartosS6calib} and~\cite{2014vircal} report uncertainty on $h(t)$ for LIGO and Virgo detectors. Additionally, an uncertainty on the scaling of $h(t)$ was reported in~\cite{2011bartosS6calib} and should be treated as a systematic error similar to the systematic waveform uncertainties discussed below that could over- or under-bias the amplitude of a signal. However, the uncertainty on the scaling of h(t) also has an associated random error that we fold into the random uncertainty calculation for Period~2. We find an overall distance uncertainty of 14\% corresponding to a 42\% uncertainty on the visible volume for Period~2. See~\cite{2013rdcalibration} for a detailed explanation of how the uncertainties were propagated.

In addition to the systematic error associated with the overall scaling of $h(t)$ that could lead to amplitude bias as mentioned above, there is a larger source of systematic error due to differences between the injected model waveforms and the true waveform. For EOBNRv2 waveforms below $\sim250\,$M$_\odot$, comparisons with numerical models indicate that uncertainties in these waveforms result in $\le10$\% systematic uncertainty in the SNR, corresponding to a $\le30$\% uncertainty in sensitive volume. For higher masses, the systematic uncertainty in the SNR could be as high as 25\%. Due to our incomplete knowledge of the true waveform and its changing uncertainty over the mass range we have explored, no systematic errors associated with imperfect waveform modeling were applied to the rate upper limits reported in this paper. Systematic errors were also not applied to previous searches~\cite{2011abadieS5highmass,2013aasiS6highmass} using full coalescence waveforms up to $100\,$M$_\odot$ and thus we can compare the upper limits directly with those results. A previous weakly modeled burst search~\cite{2012s5imbh} used waveform errors of $\sim\,\,$15\%. Thus, in order to compare with these results, the upper limits reported here should be rescaled as described below. Regarding ringdown waveforms, due to our lack of knowledge about the population of black holes producing the waveforms and the waveforms themselves, we again assign no systematic error to rate upper limits computed with ringdown waveforms.

In general, we can rescale our rate upper limits by any systematic uncertainty by applying the scaling
factor $(1-\sigma)^{-3}$ where $\sigma$ is the systematic uncertainty. Thus, we can apply
a conservative systematic uncertainty of 15\% by rescaling our rate upper limit upward by a
factor of 1.63.

The statistical error originating from the finite number of Monte Carlo injections that we have performed is the final source of error for which we must account. These errors on the efficiency at a given distance are found to range between 1.7\% and 6.2\% and were marginalized over using the method described in~\cite{2009biswasloudest, 2013corrigendum}. 

\subsection{Rate limits from full coalescence injections}
\label{sec:eobul}
In order to evaluate the sensitivity of the ringdown search to waveforms from binary IMBH coalescing systems with non-spinning components, we used a set of injections from the EOBNRv2 waveform family described in Section~\ref{sec:sim}. Due to the variation in ringdown search sensitivity over different mass ratios, we chose to compute IMBH coalescence rate upper limits separately for $q=1$ and $q=4$. The injection sets were distributed uniformly over a total binary mass range from $50 \le M/\mathrm{M}_\odot \le 450$ and upper limits were computed in mass bins of width $50\,$M$_\odot$. The final black hole spins of these injections can be determined from the mass ratios and zero initial component spins~\cite{2009finalspin}. For $q=1$, we find $\hat{a}=0.69$, and for $q=4$, we find $\hat{a}=0.47$.

\begin{figure}
\centerline{\includegraphics[width=0.5\textwidth]{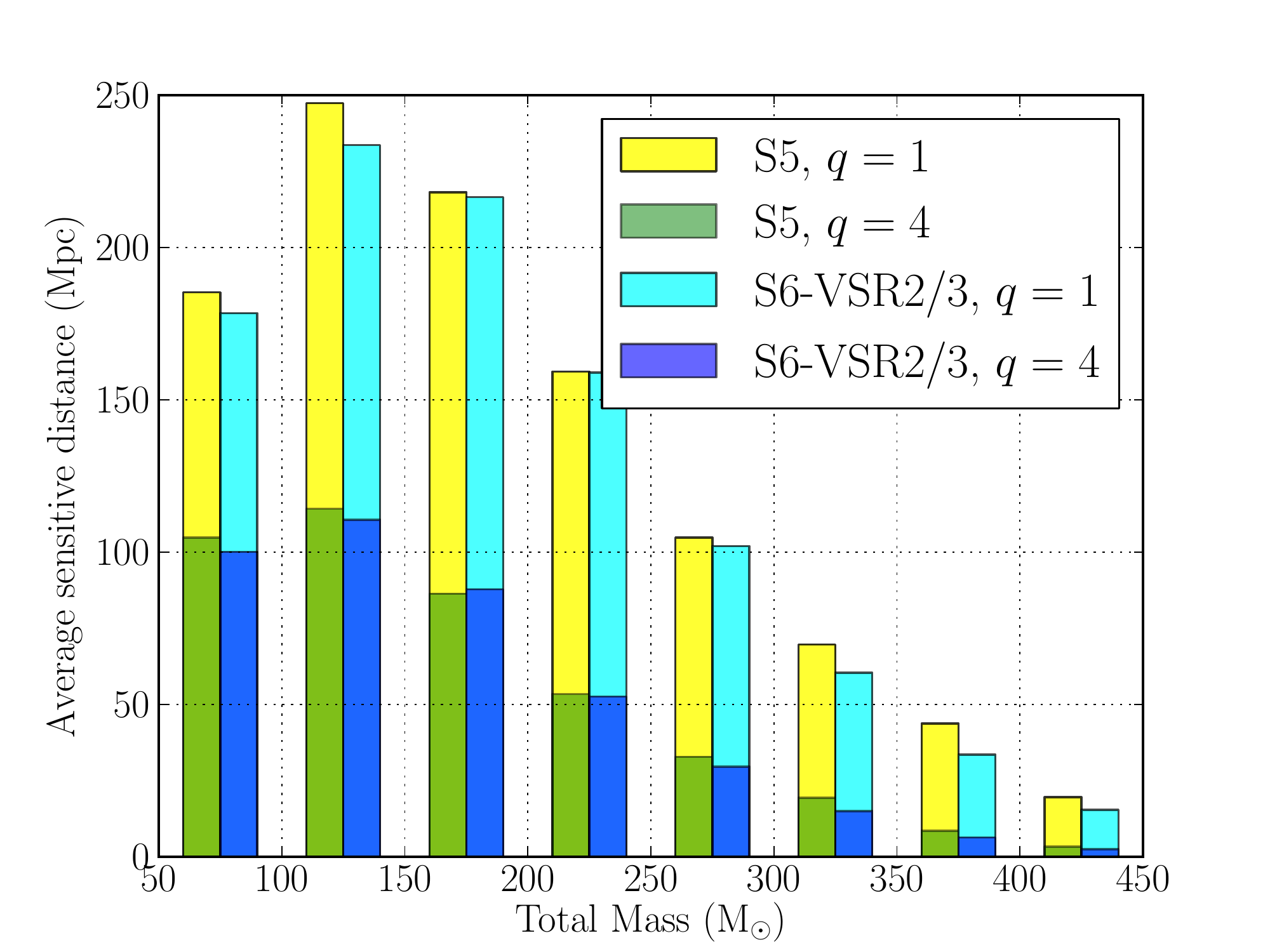}}
  \caption{Average sensitive distances of the ringdown search to binary systems described by EOBNRv2 signal waveforms over a range of total binary masses for Period~1 [$q=1$ ({\it yellow}), $q=4$ ({\it green})] and Period~2 [$q=1$ ({\it cyan}), $q=4$ ({\it blue})]. These distances are equivalent to appropriate averages over each of the detector networks
shown for Period~1 and Period~2 in Table~\ref{tab:livetime}, weighted by the percentage of time analyzed for each network. Thus, while in general the H1L1V1 and H1L1 networks during Period~2 were more sensitive than the H1H2L1 and H1L1 networks during Period~1, the consistently smaller average sensitive distances for Period~2 reflect the large duty cycle of its least sensitive detector networks compared to Period~1.}\label{fig:s5s6sensitivedist}
\end{figure}

The average sensitive distances of the ringdown search to IMBH binaries described by EOBNRv2 signal waveforms for both $q=1$ and $q=4$ are shown in Fig.~\ref{fig:s5s6sensitivedist} for Period~1 and Period~2. The most sensitive mass bin in both cases is $100\le M/\mathrm{M}_\odot \le150$ corresponding roughly to $110 \le f_0/\mathrm{Hz} \le 170$ near the peak sensitivity of the LIGO detectors. For $q=1$, the average sensitive distance of the $100 \le M/\mathrm{M}_\odot \le 150$ mass bin was $240\,$Mpc. For $q=4$, the average sensitive distance for this mass bin decreases by more than a factor of two to $110\,$Mpc. As discussed in Section~\ref{sec:waveform}, the reduced ringdown efficiency for $q=4$ binary systems leads to lower amplitude waveforms and hence, to lower average sensitive distances. Additionally, the lower final black hole spin for $q=4$ binary systems acts to decrease the average sensitive distance relative to $q=1$ binary systems for which the final spin is larger. The sensitive distance of higher mass bins drops off significantly due to the steeply rising seismic noise in the detector at low frequencies. This affect is accentuated for $q=4$ systems relative to $q=1$ systems at a fixed mass because a smaller final spin leads to a lower frequency ringdown. The sensitive distance of mass bin $400 \le M/\mathrm{M}_\odot \le 450$ is over an order of magnitude less than the sensitive distance of our most sensitive mass bins for both $q=1$ and $q=4$ cases.

\begin{figure}
\centerline{\includegraphics[width=0.5\textwidth]{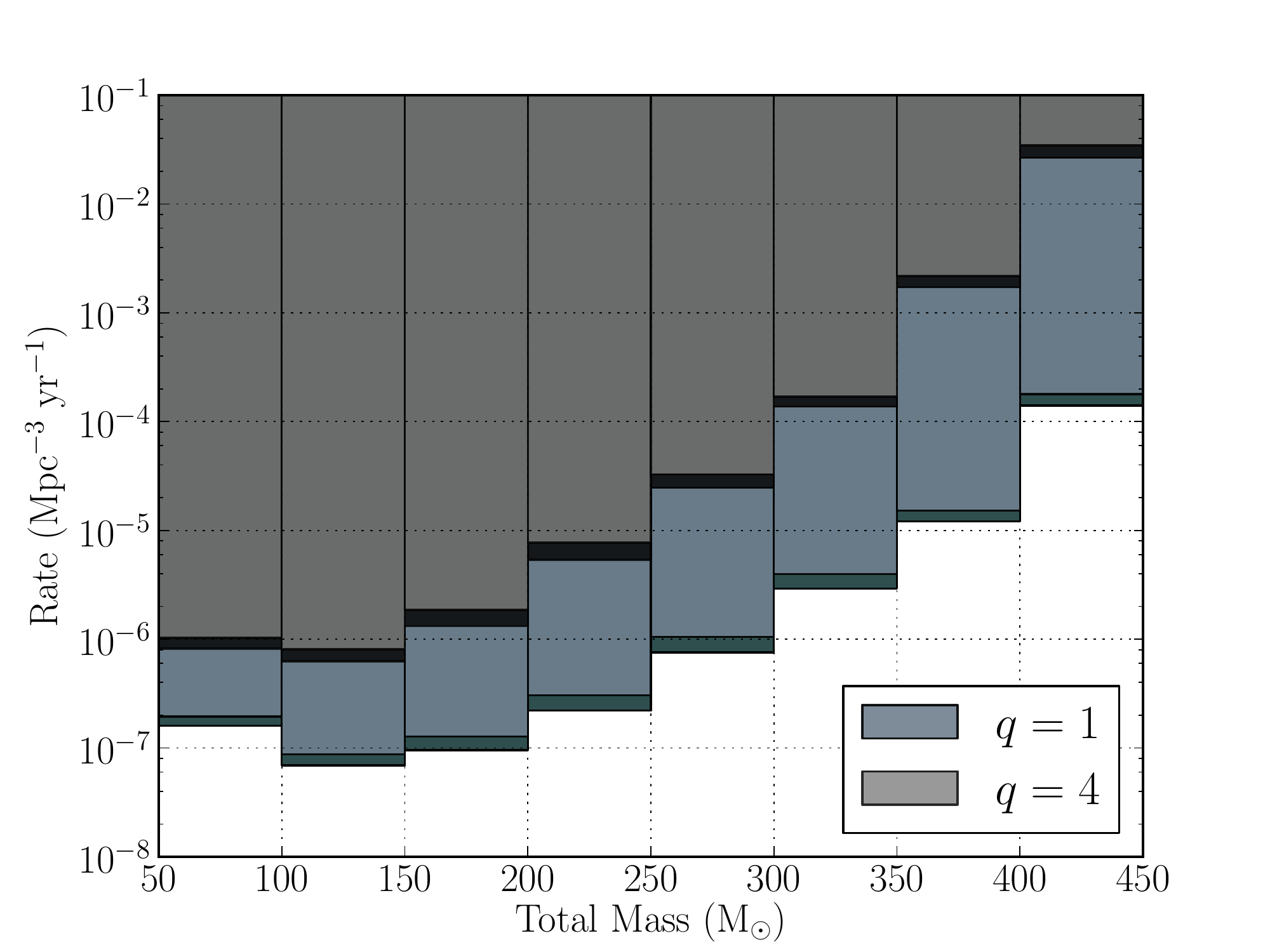}}
  \caption{Upper limits (90\% confidence) on IMBH coalescence rate in units of Mpc$^{-3}$yr$^{-1}$ as a function of total binary masses, evaluated using EOBNRv2 waveforms with $q=1$ ({\it slate grey}) and $q=4$ ({\it grey}). In both cases, upper limits computed using Period~2 with Period~1 as a prior are shown in a darker shade. Overlaid in a lighter shade are upper limits computed using only Period~1 data with a uniform prior on rate.}\label{fig:s5s6ul}
\end{figure}

Figure~\ref{fig:s5s6ul} shows the 90\%-confidence upper limits on non-spinning IMBH coalescence rates for a number of mass bins. We find an upper limit of 0.069$\times10^{-6}\,$Mpc$^{-3}$ yr$^{-1}$ on the coalescence rate of equal mass IMBH binaries with non-spinning components and total masses $100 \le M/\mathrm{M}_\odot \le 150$. From the discussion of astrophysical rates of IMBH mergers in Section~\ref{sec:source}, we see that this rate upper limit is still several orders of magnitude away from constraining the astrophysical rate from GCs.

Previous searches for weakly-modeled burst signals found no plausible events~\cite{2012s5imbh, 2013imbhs6}. The most recent search reports a rate upper limit for non-spinning IMBH coalescences of $0.12\times10^{-6}\,$Mpc$^{-3}$yr$^{-1}$ at the 90\%-confidence level for the mass bin centered on $m_1=m_2=88\,$M$_\odot$~\cite{2013imbhs6}. A direct comparison of our $q=1$ upper limits shown in Fig.~\ref{fig:s5s6ul} to this burst search result should be made with care due to the following differences between the two analyses: statistical approaches leading to different search thresholds, treatment of uncertainties, analyzed detector networks, and mass and distance binnings. Additionally, while the ringdown search employed the Bayesian formulation~\cite{2009biswasloudest, 2013corrigendum} for calculating the rate upper limit, the burst search used a frequentist method. Nevertheless, although the impact of the reported differences is hard to quantify, the upper limits determined by the two analyses can be considered consistent with each other. A more robust comparison of the sensitivity of the burst searches and an earlier version of the ringdown search without a multivariate classifier will be presented in a future paper~\cite{2013imr}.

Additionally, we can make a comparison with the upper limits reported from the matched filter search for gravitational waves from the inspiral, merger, and ringdown of non-spinning binary black holes with total masses $25 \le M/\mathrm{M}_\odot \le 100$~\cite{2013aasiS6highmass}. This search considered similar uncertainties and similar analyzed networks to those used by the ringdown search so  a result comparison is fairly straight-forward. From Table~I of~\cite{2013aasiS6highmass}, we find that for systems with $q=1$, the rate upper limits for masses $46\,\mathrm{M}_\odot$ to $100\,\mathrm{M}_\odot$ vary in the range $0.33\times10^{-6}\,$Mpc$^{-3} \mathrm{yr}^{-1}$ to $0.070\times10^{-6}\,$Mpc$^{-3} \mathrm{yr}^{-1}$. From Fig.~\ref{fig:s5s6ul}, we find a rate upper limit for mass bin $50 \le M/\mathrm{M}_\odot \le 100$ of $0.16\times10^{-6}\,$Mpc$^{-3} \mathrm{yr}^{-1}$, a value consistent with the BBH rate upper limit range for these masses and mass ratio.

Note that we can rescale our rate upper limits by a 15\% systematic uncertainty by applying the scaling factor of 1.63 as described in Section~\ref{sec:uncertainty}. From Fig.~\ref{fig:s5s6ul}, we find a rescaled rate upper limit of $0.11\times10^{-6}\,$Mpc$^{-3}$yr$^{-1}$ for mass bin $100\le M/\mathrm{M}_\odot \le150$ and $0.15\times10^{-6}\,$Mpc$^{-3}$yr$^{-1}$ for mass bin $150\le M/\mathrm{M}_\odot \le200$.

\subsection{Rate limits from ringdown injections}
\label{sec:rdul}
In order to compare with~\cite{2009s4ringdown}, we determined a 90\%-confidence upper limit  of $4\times10^{-8}\,$Mpc$^{-3}$yr$^{-1}$ on rates of pure ringdowns from perturbed black holes with uniformly distributed masses $85\le M/\mathrm{M}_\odot\le146$, uniformly distributed spins $0\le \hat{a} \le0.99$, and a fixed ringdown efficiency of $\epsilon=1\%$. We expect ringdown signals from IMBH mergers to emit near this efficiency in the $(\ell=m=2)$ fundamental mode if the mass ratio is near unity. However, for other sources of perturbed black holes, such as a hypermassive star collapse directly to a perturbed IMBH, we expect $\epsilon\ll1\%$. Thus, the rate upper limit reported in this section will not be applicable to such sources.

Reference~\cite{2009s4ringdown} placed a 90\% confidence upper limit on the rate of ringdowns from black holes with frequencies distributed uniformly in log$_{10}\left(f_0\right)$ in the range $70\le f_0/\mathrm{Hz}\le140$ and uniformly in quality factor $2\le Q \le20$ of $3.2\times10^{-5}\,$Mpc$^{-3}$yr$^{-1}$. Thus, a rough comparison indicates an improvement of nearly three orders of magnitude. A significant portion of this improvement results from a huge increase in the analysis time. Due to the high false alarm rate in double coincident analysis time, an upper limit was set in~\cite{2009s4ringdown} using only triple coincident time, a total of $0.0375\,$years. We analyzed both triple and double coincident time in both Period~1 and Period~2, a total of $1.2\,$years. Such an increase in analysis time results in a factor of $\sim$~32 improvement in the upper limit. Additionally, a significant improvement in detector sensitivity due to detector upgrades between science runs contributed to a better upper limit. Furthermore, since only triple coincident time was analyzed in~\cite{2009s4ringdown}, the sensitivity was limited by the least sensitive detector, H2, which was shown to have a horizon distance of $\sim$~$130\,$Mpc at $250\,$M$_\odot$ as shown in Fig.~2 in~\cite{2009s4ringdown}. However, since we analyzed both triple and double coincident triggers, the limiting detector was typically the L1 detector. We can compare the H2 horizon distance in Fig.~2 in~\cite{2009s4ringdown} to the L1 horizon distance in Fig.~\ref{fig:sensitivity} at $250\,$M$_\odot$ to see that the horizon distance of the limiting detector improved by a factor of $\sim$~3 for $\hat{a}=0.9$. Since the upper limit scales with volume, a factor of $\sim$~3 in distance results in a factor of $\sim$~27 in the upper limit. However, we expect this factor of improvement to decrease for the lower masses on which the ringdown upper limit was set.

Thus, from the improvements both in analysis time and detector sensitivity, we find already roughly three orders of magnitude improvement. However, several caveats would apply to a direct comparison: different injection distributions in $(M, \hat{a})$-space, the improvements from pipeline enhancements such as the implementation of a machine-learning algorithm, differences in the fitting functions for final black hole mass and spin defined in Eq.~(\ref{eq:f0}) and~(\ref{eq:q}), differences in the method used in the volume integral in Eq.~(\ref{eq:veff}), and differences in marginalization over errors. A careful study of the improvement due to the use of a machine-learning algorithm will be presented in a future paper.

\section{Summary}
\label{sec:summary}
This paper presents the results of the search for ringdown gravitational waves in data collected by LIGO and Virgo between 2005 and 2010. No significant gravitational wave candidate was identified. We place upper limits on the merger rates of non-spinning IMBH binaries as well as on the rates of ringdowns from perturbed black holes.

We conducted a detailed study of the pipeline's sensitivity to full coalescence IMBH merger signals using non-spinning EOBNRv2 waveforms. For simplicity, we focused our studies on only two mass ratios: $q=1$ and $q=4$. The average sensitive distances in our most sensitive total mass bin, $100 \le M/\mathrm{M}_\odot \le 150$, indicate that the ringdown search is sensitive to an equal mass system at twice the distance of a 4:1 mass ratio system. The most efficiently detected mass bin gives an upper limit on the rate of non-spinning, equal mass IMBH mergers with total masses $100 \le M/\mathrm{M}_\odot \le 150$ of $6.9\times10^{-8}\,$Mpc$^{-3}$yr$^{-1}$. This does not account for any uncertainty in the waveform, which could be as high as 10\% for the mass bin. Our upper limits for ringdown waveforms from perturbed IMBHs with masses $85 \le M/\mathrm{M}_\odot \le 146$ and spins $0\le \hat{a} \le0.99$ show an improvement of nearly three orders of magnitude over the previous result reported~\cite{2009s4ringdown}, which we can attribute to improved detector sensitivity, increased livetime, and pipeline enhancements.

While our rate upper limits are still two to three orders of magnitude away from constraining the astrophysical IMBH merger rate from globular clusters, we note that we will soon approach this optimistic rate with the improved sensitivity of Advanced LIGO and Virgo detectors expected to begin operation in 2015. With the improved low frequency performance of the advanced detectors, we will have sensitivity to gravitational waves from perturbed intermediate mass black holes with masses up to $\sim$ 1000 to $2000\,\mathrm{M}_\odot$. At peak sensitivity, the Advanced LIGO ringdown horizon distance for black holes with $\epsilon=1\%$ will approach cosmological distances.

\begin{center}
{\bf Acknowledgments}
\end{center}The authors gratefully acknowledge the support of the United States
National Science Foundation for the construction and operation of the
LIGO Laboratory, the Science and Technology Facilities Council of the
United Kingdom, the Max-Planck-Society, and the State of
Niedersachsen/Germany for support of the construction and operation of
the GEO600 detector, and the Italian Istituto Nazionale di Fisica
Nucleare and the French Centre National de la Recherche Scientifique
for the construction and operation of the Virgo detector. The authors
also gratefully acknowledge the support of the research by these
agencies and by the Australian Research Council, 
the International Science Linkages program of the Commonwealth of Australia,
the Council of Scientific and Industrial Research of India, 
the Istituto Nazionale di Fisica Nucleare of Italy, 
the Spanish Ministerio de Educaci\'on y Ciencia, 
the Conselleria d'Economia Hisenda i Innovaci\'o of the
Govern de les Illes Balears, the Foundation for Fundamental Research
on Matter supported by the Netherlands Organisation for Scientific Research, 
the Polish Ministry of Science and Higher Education, the FOCUS
Programme of Foundation for Polish Science,
the Royal Society, the Scottish Funding Council, the
Scottish Universities Physics Alliance, The National Aeronautics and
Space Administration, the Carnegie Trust, the Leverhulme Trust, the
David and Lucile Packard Foundation, the Research Corporation, and
the Alfred P. Sloan Foundation.

\appendix

\section{Ringdown Amplitude}
\label{sec:appendix1}
The amount of energy d$E$ carried by gravitational radiation crossing an area d$A$ orthogonal to its propagation direction in a time d$t$ is given by the energy flux equation,
\begin{equation}
\frac{\mathrm{d}E}{\mathrm{d}A \mathrm{d}t} = \frac{c^3}{16 \pi G} \left( \dot{h}_+^2 + \dot{h}_\times^2 \right),
\end{equation}
where $h_+$ and $h_\times$ are given by the generalized forms of Eq.~(\ref{eq:hplus}) and~(\ref{eq:hcross}) for an arbitrary location on a 2-sphere with $m=2$ and time of arrival $t_0$ set to zero. Taking the time derivative and squaring the plus and cross polarizations, we find
\begin{equation}
\begin{split}
\dot{h}_+^2 &= \left( \frac{\mathcal{A}}{r} \right)^2 \left( 1 + \cos^2\iota \right)^2 e^{-2\pi f_0 t/Q} \\
& \bigg[ (2\pi f_0)^2 \sin^2(2\pi f_0t + 2\phi) \\
& + \left(\frac{\pi f_0}{Q}\right)^2 \cos^2(2\pi f_0t + 2\phi)\\
& \left. + (4\pi f_0)\left(\frac{\pi f_0}{Q}\right) \sin(2\pi f_0t + 2\phi)\cos(2\pi f_0t + 2\phi) \right],
\end{split}
\end{equation}
\begin{equation}
\begin{split}
\dot{h}_\times^2 &= \left( \frac{\mathcal{A}}{r} \right)^2 \left(4\cos^2 \iota\right)e^{-2\pi f_0 t/Q} \\
& \bigg[ (2\pi f_0)^2 \cos^2(2\pi f_0t + 2\phi) \\
& + \left(\frac{\pi f_0}{Q}\right)^2 \sin^2(2\pi f_0t + 2\phi) \\
& \left. -  (4\pi f_0)\left(\frac{\pi f_0}{Q}\right) \cos(2\pi f_0t + 2\phi)\sin(2\pi f_0t + 2\phi) \right] .
\end{split}
\end{equation}
Integrating this flux over a sphere with area element d$A=r^2\mathrm{d}(\cos\iota)\mathrm{d}\phi$, we find that the trigonometric functions simplify greatly, leaving only the exponential time dependence over which to integrate
\begin{equation}
\begin{split}\label{eq:denergy}
E &= \frac{c^3}{16 \pi G} \int_T \int_A \left( \dot{h}_+^2 + \dot{h}_\times^2 \right) \mathrm{d}A \mathrm{d}t\\
&=  \frac{8 c^3}{5 G} \mathcal{A}^2 (\pi^2 f_0^2)\left( 1 + \frac{1}{4Q^2} \right) \int_{t=0}^{\infty} e^{-2\pi f_0 t/Q} \mathrm{d}t \\
&= \frac{4c^3}{5 G}\mathcal{A}^2 (\pi f_0) \left( 1 + \frac{1}{4Q^2} \right) Q.
\end{split}
\end{equation}
Finally, we note that the energy radiated as gravitational waves during the ringdown phase is $E=\epsilon Mc^2$ where $\epsilon$ is the ringdown efficiency discussed in Sec.~\ref{sec:waveform}. Thus, the amplitude can be found by solving Eq.~(\ref{eq:denergy}) for $\mathcal{A}$,
\begin{equation}
\begin{split}\label{eq:rdamp}
\mathcal{A} = \sqrt{\frac{5\epsilon GM}{4\pi c}} f_0^{-1/2}\left( 1 + \frac{1}{4Q^2} \right)^{-1/2} Q^{-1/2}.
\end{split}
\end{equation}

\section{Ringdown Horizon Distance}
\label{sec:appendix2}
The ringdown horizon distance, similar to the inspiral horizon distance, is a useful measure of the sensitivity of the detectors to ringdown gravitational waves from a particular type of black hole. It is equal to the distance at which an optimally oriented and located IMBH merger would produce an SNR of 8 in the detector. The horizon distance is derived from the representative strain noise power spectral density of a detector and the $h_\mathrm{rss}$, or root sum squared of the strain, for a signal with optimal orientation at $1\,$Mpc. The definition of $h_\mathrm{rss}$ comes from the need to measure the amplitude of a gravitational wave without reference to a particular detector. In general, it is
\begin{equation}
h_\mathrm{rss}^2=\int_0^{\infty} \left( h_+^2(t) + h_\times^2(t) \right) \mathrm{d}t,
\end{equation}
where $h_+$ and $h_\times$ are given in Eq.~(\ref{eq:hplus}) and~(\ref{eq:hcross}) for the single-mode $(\ell, m, n) = (2,2,0)$ ringdown waveform. Here, under the assumption of optimal orientation, we set $\iota=0$. We find that the $h_\mathrm{rss}$ takes the form
\begin{equation}\label{eq:hrsssq}
h_\mathrm{rss}^2=4\left(  \frac{\mathcal{A}}{r} \right)^2 \left( \frac{Q}{2\pi f_0} \right).
\end{equation}
where $\mathcal{A}$ is derived in Eq.~(\ref{eq:rdamp}). If $\tilde{h}(f)$ represents the Fourier transform of the expected signal, then the average SNR this signal would attain in a detector with spectral density $S_n(f)$ is given by
\begin{equation}\label{eq:rho1}
\left<\rho\right> = \sqrt{4 \int_0^{\infty} \frac{\left| \tilde{h}(f)\right|^2}{S_n(f)} df}.
\end{equation}
Typically, the horizon distance is found by setting $\left<\rho\right>=8$ and solving for the distance $r$ which parameterizes the waveform $\tilde{h}$. We can use the fact that the single-mode ringdown signal is quasi-monochromatic and $S_n(f)$ assumes approximately one value for each $f_0$ so it can be treated as a constant:
\begin{equation}\label{eq:rho2}
\left<\rho\right> =  \sqrt{\frac{4}{S_n(f_0)} \int_{0}^{\infty} \left| \tilde{h}(f)\right|^2 df}.
\end{equation}
Using Parseval's theorem, we can write Eq.~(\ref{eq:rho2}) as
\begin{equation}\label{eq:rho3}
\left<\rho\right> = \sqrt{\frac{2}{S_n(f_0)} \int_{-\infty}^{\infty} h^2(t) dt}.
\end{equation}
Also, since optimally oriented and located sources imply maximization over all the angles $\theta$, $\phi$, and $\psi$ in $F_+$ and $F_\times$, then $F_+=1$ and $F_\times=0$. This then gives us the result that $h(t)=h_+(t)$ (which is defined for $t>0$) so Eq.~(\ref{eq:rho3}) becomes
\begin{equation}\label{eq:rho4}
\begin{split}
\left<\rho\right> &= \sqrt{\frac{2}{S_n(f_0)} \int_{0}^{\infty} h_+^2(t) dt} \\
&= \sqrt{\frac{2}{S_n(f_0)} h_\mathrm{rss}^2 \frac{1+2Q^2}{1+4Q^2}}\\
&= \sqrt{\frac{2}{S_n(f_0)} h^2_\mathrm{rss}(1\,\mathrm{Mpc})\left(\frac{1\,\mathrm{Mpc}}{r}\right)^2 \frac{1+2Q^2}{1+4Q^2}}
\end{split}
\end{equation}
where $h^2_\mathrm{rss}(1\,\mathrm{Mpc})$ is Eq.~(\ref{eq:hrsssq}) evaluated at a distance of $1\,$Mpc. Then, we simply solve Eq.~(\ref{eq:rho4}) for the horizon distance,
\begin{equation}
r = \frac{1\,\mathrm{Mpc}}{\left<\rho\right>} \sqrt{\frac{2}{S_n(f_0)}h^2_\mathrm{rss}(1\,\mathrm{Mpc}) \frac{1+2Q^2}{1+4Q^2}}.
\end{equation}
We then set $\left<\rho\right>=8$ to define the ringdown horizon distance used in Fig.~\ref{fig:sensitivity}.

\end{document}